\documentclass[print]{aa}  
\usepackage{amssymb,amsmath}
\usepackage{graphicx}
\usepackage{txfonts}
\usepackage{natbib}
\usepackage{pslatex}
\usepackage{caption}
\usepackage{changes}
\usepackage{longtable}
\usepackage{subfigure}
\usepackage{lscape}
\usepackage{threeparttable}

\usepackage{color}
\usepackage{amsmath}
\begin{document}

   \title{Evidence of a decreasing trend for the Hubble constant}

   \author{X. D. Jia
          \inst{1}
          \and
          J. P. Hu\inst{1}
          \and
          F. Y. Wang\inst{1,2}\thanks{E-mail: fayinwang@nju.edu.cn}
          }

   \institute{School of Astronomy and Space Science, Nanjing University, Nanjing 210093, China\\
         \and
             Key Laboratory of Modern Astronomy and Astrophysics (Nanjing University), Ministry of Education, Nanjing 210093, China\\
             }


 
 \abstract{
  The current discrepancy between the Hubble constant, $H_0$, derived from the local distance ladder and from the cosmic microwave background is one of the most crucial issues in cosmology, as it may possibly indicate unknown systematics or new physics. Here, we present a novel non-parametric method to estimate the Hubble constant as a function of redshift. We establish independent estimates of the evolution of Hubble constant by diagonalizing the covariance matrix. From type Ia supernovae and the observed Hubble parameter data, a decreasing trend in the Hubble constant with a significance of a 5.6$\sigma$ confidence level is found. At low redshift, its value is dramatically consistent with that measured from the local distance ladder and it drops to the value measured from the cosmic microwave background at high redshift. Our results may relieve the Hubble tension, with a  preference for recent solutions, especially with respect to  novel physics.}
  

\maketitle
%

\section{Introduction}\label{sec:intro}
The standard cosmological-constant $\Lambda$ cold dark matter (CDM) model has been widely accepted and shown to be remarkably successful in explaining
most cosmological observations  \citep{2020A&A...641A...6P,2021PhRvD.103h3533A}. 
However, the standard $\Lambda$CDM model is challenged by tensions among
different probes \citep{Perivolaropoulos2022}. The most serious one is the Hubble tension \citep{Verde2019,Riess2020}. The extrapolation from fitting the $\Lambda$CDM model to the $Planck$ cosmic microwave background (CMB) anisotropies measurements gives the Hubble constant $H_{0,z\sim1100} = 67.4\pm 0.5$ km s$^{-1}$ Mpc$^{-1}$ \citep{2020A&A...641A...6P}. Using the local distance ladder, such as Cepheids and type Ia supernovae (SNe Ia), the SH0ES team measured $H_{0,z\sim0} = 73.04\pm 1.04$ km s$^{-1}$ Mpc$^{-1}$ \citep{2022ApJ...934L...7R}. The discrepancy between the two measurements is about 5$\sigma$ \citep{2022ApJ...934L...7R}. 
The essence of Hubble tension is that the values of $H_0$ derived from cosmic observations at different redshifts are inconsistent.
If there is no evidence for systematic uncertainties in the data analysis, the tension may indicate a defect of the standard $\Lambda$CDM model.

To solve the Hubble tension, some theoretical models have been proposed \citep{2021CQGra..38o3001D,2021A&ARv..29....9S} and can be divided into two broad classes. One is aimed at proposing changes to the late-time universe, as in the case of some modified gravity theories \citep{Haslbauer2020,2021MNRAS.500.1795B}, local inhomogeneity theory \citep{Marra2013,2019ApJ...875..145K}, dark matter models \citep{2022arXiv220908102N}, and dark energy models \citep{2017NatAs...1..627Z,2021PhRvD.103l1302C} have been extensively studied to relieve the Hubble tension. In contrast, early Universe resolutions, which modify pre-recombination physics, have focused on dark radiation \citep{2016JCAP...10..019B}, strong
neutrino self-interactions \citep{Kreisch2020}, and early dark energy \citep{2019PhRvL.122v1301P,Sakstein2020,2021PhRvD.104f3524V,2022arXiv220807631R}. However, none of the proposed models can resolve the Hubble tension successfully \citep{2022PhR...984....1S}.

Theoretically, the value of Hubble constant -- $H_{0,z}$, defined as the value of $H_0$ derived from the cosmic observations at redshift, $z$, that is, the value of $H_{0,z\sim1100} = 67.4\pm 0.5$ km s$^{-1}$ Mpc$^{-1}$  derived from {\it Planck} CMB measurements at $z\sim 1100$ -- may be redshift-dependent. There are several plausible reasons for this. First, in the Friedmann-Lema\^itre-Robertson-Walker (FLRW) framework, we can obtain 
$H_{0,z}=H(z)\exp\left({-3\int_0^{z}(1+w_{\rm eff}(z'))/(1+z')/2dz'}\right)$
by integrating the Friedmann equations, where $w_{\rm eff}(z)=\Sigma_i^n \Omega_iw_i$ for a cosmological model consisting of $n$ component
fluid with energy densities, $\rho_i$, and equation of state, $w_i$ {\citep{2022arXiv220113384K}.
So, the value of $H_{0,z}$ is determined by extrapolating the $H(z)$ (or luminosity distance $d_L$) from observational data at
higher $z$ to $z=0$ after assuming a cosmological model. If the effective equation of state (EoS) $w_{\rm eff}$ varies with redshift, the derived $H_{0,z}$ may be not a constant value. It is only under the assumption that $H_{0,z}$ is a constant that the effect of $w_{\rm eff}(z)$'s evolution be compensated by $H(z)$. When we use the observational data to estimate the value of $H_{0,z}$, the behavior of $w_{\rm eff}$ will make a difference to $H_{0,z}$.  
Second, evidence for local voids has been varied among studies using galaxy catalogs \citep{2022MNRAS.511.5742W}. This inhomogeneity will lead to the evolution of $H_{0,z}$ with redshifts. Third, some modified gravity models can explain the late-time cosmic acceleration \citep{2002IJMPD..11..483C} and it may also lead to the evolution of $H_{0,z}$ \citep{Kazantzidis2020,2021ApJ...912..150D}. 

Recently, some marginal evidence has shown that the value of Hubble constant may evolve with redshift \citep{Kazantzidis2020,Hu2022,universe9020094}. In the flat $\Lambda$CDM model, a descending trend in $H_{0,z}$ with redshift has been found \citep{2020PhRvD.102j3525K,Wong2020,2021ApJ...912..150D,2022Galax..10...24D,2022PhRvD.106d1301O,2022arXiv220611447C,2023arXiv230112725M}. The evolution of Hubble constant may be a potential solution of the Hubble tension. However, there is a degeneracy between the derived $H_{0,z}$ at different redshifts (with more details given in Sect. \ref{Sec3}). Here, we present a novel non-parametric method to estimate Hubble constant at different redshifts. 

In this work, we investigate the redshift-evolution of $H_{0,z}$ with a non-parametric approach.
In this scenario, $H_{0,z}$ is not a constant, but a piecewise function with redshift. We allow
$H_{0,z_i}$ to be a constant value in redshift bin $i$. For a given observation, the $H_{0,z_i}$ values will generally be correlated with each other because measurements of luminosity distance, $d_L$, and Hubble parameter, $H(z),$ constrain redshift summations of $H_{0,z_i}$, which is similar to estimating the dark energy EoS as a function of redshift \citep{2005PhRvD..71b3506H,Riess2007,Jia2022}. The degeneracy among $H_{0,z}$ is removed by diagonalizing the covariance matrix.

\section{The data sample}\label{Sec2}
In this paper, we use the latest observational data to constrain the cosmological parameters. The joint data sample contains $H(z)$ measurements, baryon acoustic oscillation (BAO) data, and the SNe Ia sample.
\linespread{1}
\begin{table}
    \caption{Expansion data from cosmic chronometers. \label{CCsample}}
    \centering
    \begin{tabular}{ccc} 
    \hline
    $z$  & $H(z)$ $\textrm{km}~\textrm{s}^{-1} \textrm{Mpc}^{-1}$   & Ref.   \\ \hline
0.07 & 69.0 $\pm$ 19.6 & \cite{2014RAA....14.1221Z} \\  \hline 
0.09 & 69.0 $\pm$ 12.0 & \cite{2003ApJ...593..622J} \\  \hline 
0.12 & 68.6 $\pm$ 26.2 & \cite{2014RAA....14.1221Z} \\  \hline 
0.17 & 83.0 $\pm$ 8.0 & \cite{2005PhRvD..71l3001S} \\  \hline 
0.179 & 75.0 $\pm$ 4.0 & \cite{2012JCAP...08..006M} \\  \hline 
0.199 & 75.0 $\pm$ 5.0 & \cite{2012JCAP...08..006M} \\  \hline 
0.2 & 72.9 $\pm$ 29.6 & \cite{2014RAA....14.1221Z} \\  \hline 
0.27 & 77.0 $\pm$ 14.0 & \cite{2005PhRvD..71l3001S} \\  \hline 
0.28 & 88.8 $\pm$ 36.6 & \cite{2014RAA....14.1221Z} \\  \hline 
0.352 & 83.0 $\pm$ 14.0 & \cite{2012JCAP...08..006M} \\  \hline 
0.38 & 83.0 $\pm$ 13.5 & \cite{2016JCAP...05..014M} \\  \hline 
0.4 & 95.0 $\pm$ 17.0 & \cite{2005PhRvD..71l3001S} \\  \hline 
0.4 & 77.0 $\pm$ 10.2 & \cite{2016JCAP...05..014M} \\  \hline 
0.425 & 87.1 $\pm$ 11.2 & \cite{2016JCAP...05..014M} \\  \hline 
0.45 & 92.8 $\pm$ 12.9 & \cite{2016JCAP...05..014M} \\  \hline 
0.47 & 89.0 $\pm$ 34.0 & \cite{2017MNRAS.467.3239R} \\  \hline 
0.478 & 80.9 $\pm$ 9.0 & \cite{2016JCAP...05..014M} \\  \hline 
0.48 & 97.0 $\pm$ 62.0 & \cite{2010JCAP...02..008S} \\  \hline 
0.593 & 104.0 $\pm$ 13.0 & \cite{2012JCAP...08..006M} \\  \hline 
0.68 & 92.0 $\pm$ 8.0 & \cite{2012JCAP...08..006M} \\  \hline 
0.75 & 98.8 $\pm$ 33.6 & \cite{2022ApJ...928L...4B} \\  \hline 
0.781 & 105.0 $\pm$ 12.0 & \cite{2012JCAP...08..006M} \\  \hline 
0.8 & 113.1 $\pm$ 28.5 & \cite{2022arXiv220505701J} \\  \hline 
0.875 & 125.0 $\pm$ 17.0 & \cite{2012JCAP...08..006M} \\  \hline 
0.88 & 90.0 $\pm$ 40.0 & \cite{2010JCAP...02..008S} \\  \hline 
0.9 & 117.0 $\pm$ 23.0 & \cite{2005PhRvD..71l3001S} \\  \hline 
1.037 & 154.0 $\pm$ 20.0 & \cite{2012JCAP...08..006M} \\  \hline 
1.3 & 168.0 $\pm$ 17.0 & \cite{2005PhRvD..71l3001S} \\  \hline 
1.363 & 160.0 $\pm$ 33.6 & \cite{2015MNRAS.450L..16M} \\  \hline 
1.43 & 177.0 $\pm$ 18.0 & \cite{2005PhRvD..71l3001S} \\  \hline 
1.53 & 140.0 $\pm$ 14.0 & \cite{2005PhRvD..71l3001S} \\  \hline 
1.75 & 202.0 $\pm$ 40.0 & \cite{2005PhRvD..71l3001S} \\  \hline 
1.965 & 186.5 $\pm$ 50.4 & \cite{2015MNRAS.450L..16M} \\  \hline 
                \hline
        \end{tabular}~~~~~~~~
\end{table}
The Hubble parameter sample contains 33 $H(z)$ measurements \citep{2018ApJ...856....3Y,2022MNRAS.513.5686C}, covering redshifts from 0.07 to 1.965. These 33 $H(z)$ data are derived using the cosmic chronometic technique.
The data are shown in Table \ref{CCsample}. This method compares the differential age evolution of galaxies that are at different redshifts by $H(z) = -\frac{1}{1+z} \frac{d z}{d t}$ \citep{2002ApJ...573...37J}.
The value of $dz/dt$ can be approximately replaced by $\Delta z$/$\Delta t$, where $\Delta t$ is the measurements of the age difference between two passively evolving galaxies and $\Delta z$ is the small redshift interval between them. 

\linespread{1}
\begin{table*}
\begin{threeparttable}
    \caption{BAO data. \label{BAOsample}}
    \centering
    \begin{tabular}{cccc} 
    \hline
    $z$ & parameter\tnote{a} & value &  ref.   \\ \hline
    $0.122$ & $D_V\left(r_{s,{\rm fid}}/r_s\right)$ & $539\pm17$ & \cite{2018MNRAS.481.2371C}\\
    \hline
    $0.38$ & $D_M/r_s$ & 10.23406 & \cite{2020MNRAS.498.2492G}\tnote{b}\\
    \hline
    $0.38$ & $D_H/r_s$ & 24.98058 & \cite{2020MNRAS.498.2492G}\tnote{b}\\
    \hline
    $0.51$ & $D_M/r_s$ & 13.36595 & \cite{2020MNRAS.498.2492G}\tnote{b}\\
    \hline
    $0.51$ & $D_H/r_s$ & 22.31656 & \cite{2020MNRAS.498.2492G}\tnote{b}\\
    \hline
    $0.698$ & $D_M/r_s$ & 17.85823691865007 & \cite{2020MNRAS.498.2492G,2021MNRAS.500..736B}\tnote{c}\\
    \hline
    $0.698$ & $D_H/r_s$ & 19.32575373059217 & \cite{2020MNRAS.498.2492G,2021MNRAS.500..736B}\tnote{c}\\
    \hline
    $0.81$ & $D_A/r_s$ & $10.75\pm0.43$ & \cite{2019MNRAS.483.4866A}\\
    \hline
    $1.48$ & $D_M/r_s$ & 30.6876 & \cite{2020MNRAS.499..210N,2021MNRAS.500.1201H}\tnote{d}\\
    \hline
    $1.48$ & $D_H/r_s$ & 13.2609 & \cite{2020MNRAS.499..210N,2021MNRAS.500.1201H}\tnote{d}\\
    \hline
    $2.334$ & $D_M/r_s$ & 37.5 & \cite{2020ApJ...901..153D}\tnote{e}\\
    \hline
    $2.334$ & $D_H/r_s$ & 8.99 & \cite{2020ApJ...901..153D}\tnote{e}\\
    \hline\hline
        \end{tabular}~~~~~~~~
\begin{tablenotes}[flushleft]
\item[a] $D_V$, $r_s$, $r_{s, {\rm fid}}$, $D_M$, $D_H$, and $D_A$ in units of Mpc.
\item[b] The four measurements are correlated and Equation (\ref{Covb}) is their covariance matrix.
\item[c] The two measurements are correlated and Equation (\ref{Covc}) is their covariance matrix.
\item[d] The two measurements are correlated and Equation (\ref{Covd}) is their covariance matrix.
\item[e] The two measurements are correlated and Equation \eqref{Cove} is their covariance matrix.
\end{tablenotes}
\end{threeparttable}
\end{table*}

The 12 BAO data points are shown in Table \ref{BAOsample}, spanning the redshift range $0.122 \leq z \leq 2.334 $. The covariance matrices for them are shown as follows.
The covariance matrix $\textbf{C}$ for BAO data from \cite{2020MNRAS.498.2492G} is
\begin{equation}
    \label{Covb}
    \begin{bmatrix}
    0.02860520 & -0.04939281 & 0.01489688 & -0.01387079\\
    -0.04939281 & 0.5307187 & -0.02423513 & 0.1767087\\
    0.01489688 & -0.02423513 & 0.04147534 & -0.04873962\\
    -0.01387079 & 0.1767087 & -0.04873962 & 0.3268589
    \end{bmatrix}.
\end{equation}
The BAO data from \cite{2020MNRAS.498.2492G} and \cite{2021MNRAS.500..736B} has the covariance matrix $\textbf{C}$
\begin{equation}
    \label{Covc}
    \begin{bmatrix}
    0.1076634008565565 & -0.05831820341302727\\
    -0.05831820341302727 & 0.2838176386340292 
    \end{bmatrix}.
\end{equation}
For BAO data from \cite{2020MNRAS.499..210N} and \cite{2021MNRAS.500.1201H}, the covariance matrix $\textbf{C}$ is
\begin{equation}
    \label{Covd}
    \begin{bmatrix}
    0.63731604 & 0.1706891\\
    0.1706891 & 0.30468415
    \end{bmatrix},
\end{equation}
The covariance matrix $\textbf{C}$ for BAO data from \cite{2020ApJ...901..153D} is
\begin{equation}
    \label{Cove}
    \begin{bmatrix}
    1.3225 & -0.1009 \\
    -0.1009 & 0.0380
    \end{bmatrix}.
\end{equation}
}
\par
Overall, SNe Ia are characterized a nearly uniform intrinsic luminosity and can be used as standard candles. We used the Pantheon+ SNe Ia sample, which consists of $1701$ light curves of 1550 distinct SNe Ia \citep{2021arXiv211203863S}. It includes SNe Ia that are in galaxies with measured Cepheid distances, which is important
for measuring the Hubble constant \citep{2021arXiv211203863S}.


\section{Method}\label{Sec3}
    \subsection{Non-parametric $H_{0,z}$ constraint}
The value of Hubble constant $H_{0,z}$ is measured by extrapolating the Hubble parameter, $H(z),$ or the luminosity distance, $d_L(z)$, from observational data at higher $z$ to $z=0$, by choosing a particular cosmological model. There is no prior knowledge of the redshift evolution of $H_{0,z}$. Similarly to the principal-component approach used to study the redshift evolution of the EoS of dark energy \citep{2005PhRvD..71b3506H}, 
in order to avoid adding some priors on the nature of $H_0(z)$, we do not assume that it follows some particular functions. We just allow the value of $H_{0,z}$ in each redshift bin to remain a constant. 

\begin{figure}
        \centering
        \includegraphics[width=0.5\textwidth,angle=0]{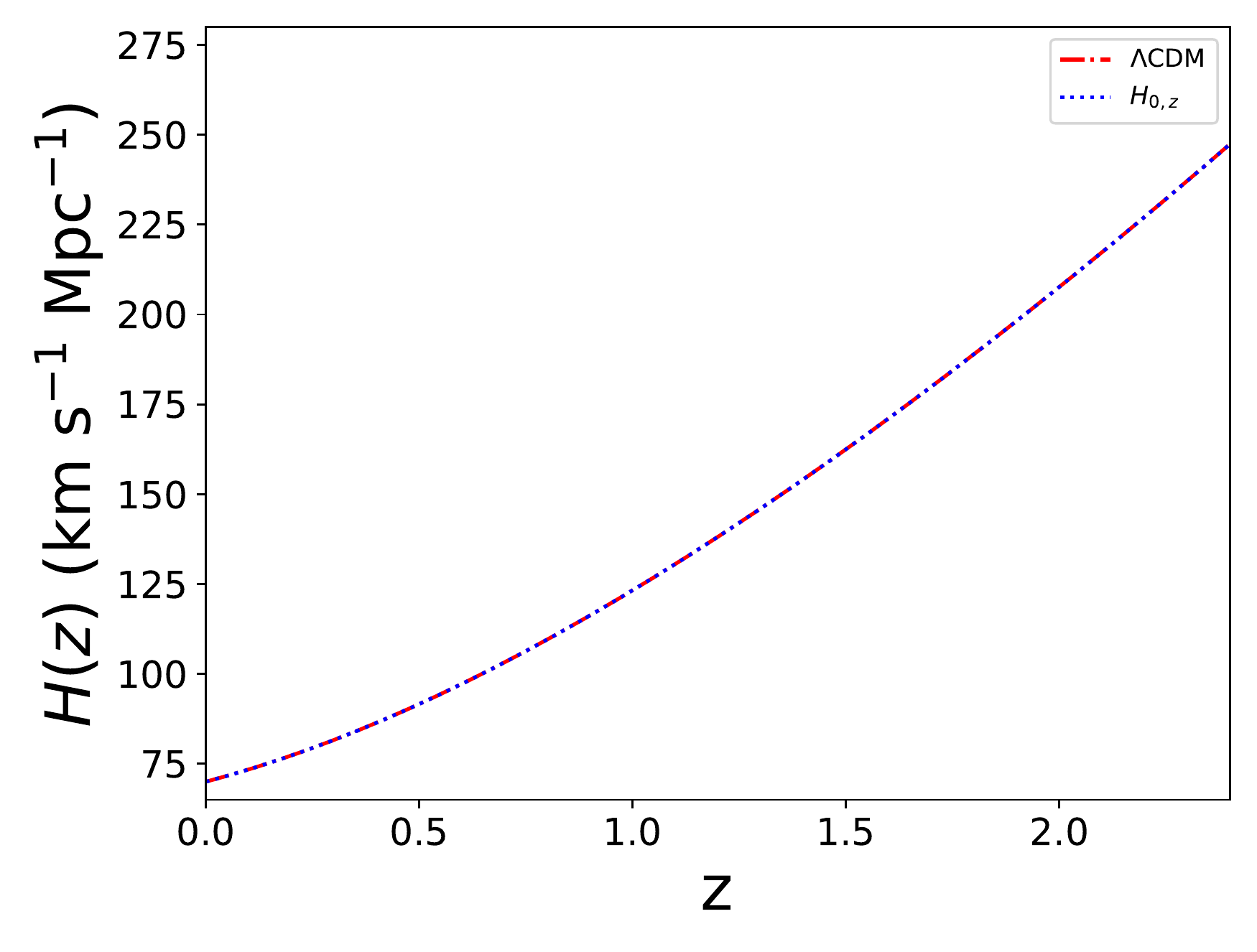}
        \caption{Comparison between $H(z)$ in standard $\Lambda$CDM and the one calculated by Equation (\ref{Hz step}). $H_0=H_{0,z_i} = 70$ km s$^{-1}$ Mpc$^{-1}$, $\Omega_{k0}=0,$ and $\Omega_{m0} = 0.3$ are assumed. }
        \label{F_Hz}
\end{figure}

Under the assumption of a piece-wise function, $H_0(z)$ can be expressed as:
\begin{equation}\label{H0function}
H_0(z)=\begin{cases} H_{0,z_1} &\text{ if } 0\le z < z_1, \\ 
H_{0,z_2} &\text{ if } z_1 \le z < z_2,\\
\cdots  &\cdots,\\
H_{0,z_i} &\text{ if } z_{i-1} \le z < z_i,\\
\cdots &\cdots, \\
H_{0,z_N} &\text{ if } z_{N-1} \le z < z_N.
\end{cases}
\end{equation}
The parameter $i$ means the $i$th redshift bin, and $N$ is the number of total redshift bins. 
As described in the previous definition of $H_{0,z}$, we use $H_{0,z_i}$ to represent the value of $H_0(z)$ between $z_{i-1}$ to $z_i$.

A simple way to model the possible evolution of $H_{0,z}$ is through a modification of the standard cosmological model.
In the $\Lambda$CDM model, the Hubble parameter is given by
\begin{equation}\label{Hz}
H(z)=H_{0} \sqrt{\Omega_{m0}(1+z)^{3}+\Omega_{k0}(1+z)^{2}+\Omega_{\Lambda0}} .
\end{equation}
The parameter $\Omega_{\mathrm{m0}}$ is the cosmic matter density, $\Omega_{\mathrm{k0}}$ is the spatial curvature and $\Omega_{\Lambda0}$ is the energy density parameter of the cosmological constant. 
According to the result from {\it Planck} CMB measurements \citep{2020A&A...641A...6P}, the space is extremely flat, which gives $\Omega_{\mathrm{k0}}=0$.

The integral form of Equation (\ref{Hz}) is
\begin{equation}\label{expansion integral}
\begin{split}
H(z)=&H_0\sqrt{\Omega_{m0}(1+z)^{3}+\Omega_{\Lambda0}} \\
=&H_0\left(\int_{0}^{z}  \frac{3\Omega_{m0}(1+z^{\prime})^2}{2\sqrt{\Omega_{m0}(1+z^{\prime})^{3}+\Omega_{\Lambda0}}} dz^{\prime}+1\right) \\
=& \int_{0}^{z}  \frac{H_03\Omega_{m0}(1+z^{\prime})^2}{2\sqrt{\Omega_{m0}(1+z^{\prime})^{3}+\Omega_{\Lambda0}}} dz^{\prime}+H_0.
\end{split}
\end{equation}
The constant 1 is determined by the equation $\Omega_{m0}+\Omega_{\Lambda0} = 1$ when $z = 0$. The function $H(z)$ in Equation (\ref{expansion integral}) is identical to that in Equation (\ref{Hz}), and it is also more convenient to segment redshift intervals. 
\par
Replacing $H_0$ in Equation (\ref{expansion integral}) with the piece-wise function in Equation (\ref{H0function}), the Hubble parameter is expressed as
\begin{equation}\label{Hz step}
\begin{split}
H(z_i)=& H_{0,z_1}\int_{0}^{z_1} \frac{3\Omega_{m0}(1+z)^2}{2\sqrt{\Omega_{m0}(1+z)^{3}+\Omega_{\Lambda0}}} \\
& +H_{0,z_2}\int_{z_1}^{z_2} \frac{3\Omega_{m0}(1+z)^2}{2\sqrt{\Omega_{m0}(1+z)^{3}+\Omega_{\Lambda0}}} \\
& +\cdots \\
& +H_{0,z_i}\int_{z_{i-1}}^{z_i} \frac{3\Omega_{m0}(1+z)^2}{2\sqrt{\Omega_{m0}(1+z)^{3}+\Omega_{\Lambda0}}}+H_{0,z_i}.
\end{split}
\end{equation}

The last term $H_0$ in Equation (\ref{expansion integral}) is replaced by $H_0(z)$, so its redshift should be $z_i$. Meanwhile, the value of $H_0(z)$ at low redshifts is the result of the evolution of high-redshift $H_0(z)$. If $H_0(z)$ does not exhibit any evolutionary trend, the result will revert to $H_0$. We show the value of $H(z)$ calculated by Equations (\ref{Hz}) and (\ref{Hz step}) in Figure \ref{F_Hz}, respectively. Here, $H_0 = H_{0,z_1} = H_{0,z_2} = \cdots = H_{0,z_i} = 70$ km s$^{-1}$ Mpc$^{-1}$, $\Omega_{k0}=0,$ and $\Omega_{m0} = 0.3$ are assumed. It is clear to see that the value of $H(z)$ calculated via Equation (\ref{Hz}) is the same as the one as in Equation (\ref{Hz step}).
Therefore, the expansion of Hubble parameter in Equation (\ref{Hz step}) is correct.

By assuming the value $H_{0,z}$ is a constant in each redshift bin, previous works have directly used Equation (\ref{Hz}) to derive $H_{0,z}$ \citep{2020PhRvD.102j3525K,2021ApJ...912..150D}. First, this approach violates the assumption that the value of $H_{0,z}$ is constant in each bin. Equation (\ref{Hz}) actually specifies that the value of $H_{0,z}$ is constant from $z=0$ to $z=z_i$ -- and not from $z=z_{i-1}$ to $z=z_i$. Therefore, the assumption leads $H_{0,z}$ as a constant from $z=0$ to each $z=z_i$. During the process of fitting the data in the $i$th redshift bin, the derived value of $H_{0}$ is in the redshift range $0-z_i$, not the value of $H_{0,z_i}$ in $z_{i-1}-z_i$. Their measured $H_{0,z_i}$ is not the value in the $i$th redshift bin. Thus, it is problematic to use the result to represent the value of $H_{0,z}$ in the $i$th redshift bin. Second, the correlation among each $H_{0,z_i}$ was not considered in previous works. Due to the fact that the Hubble parameter and luminosity distance depend on the summation and the integration over redshifts, the value of $H_{0,z_i}$ in low-redshift bins will affect the value in high-redshift bins. In our method, Equation (\ref{Hz step}) reveals the redshift-evolution of $H_0(z)$. The most obvious difference between these two methods is that Equation (\ref{Hz}) demonstrates the value of $H_{0,z}$ from $z=0$ to $z=z_i$, but Equation (\ref{Hz step})  reveals the value of $H_{0,z}$ at the $i$th redshift bin. The correlations can be removed by the principal component analysis.

\begin{figure*}
    \centering
    \includegraphics[width=\textwidth,angle=0]{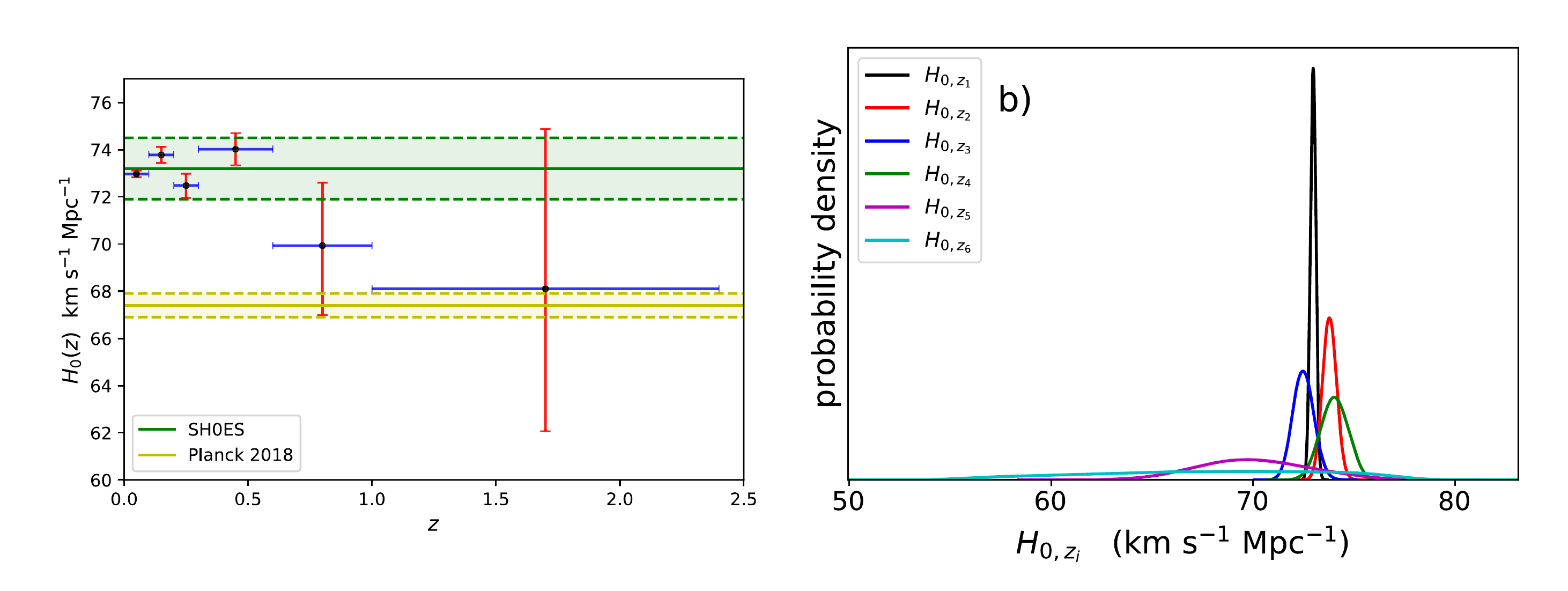}
    \caption{Fitting results of $H_0(z)$ in the Pantheon+ sample case for six redshift bins. The left panel shows the value of $H_0(z)$ as a function of redshift. There is a clear decreasing trend. The green line gives $H_{0} = 73.04\pm 1.04$ km s$^{-1}$ Mpc$^{-1}$ from local distance ladder and its 1$\sigma$ uncertainty \citep{2022ApJ...934L...7R}. The yellow line is the value of $H_{0} = 67.4\pm 0.5$ km s$^{-1}$ Mpc$^{-1}$ from CMB measurements and its 1$\sigma$ uncertainty \citep{2020A&A...641A...6P}. The right panel shows the normalized probability distributions of $H_{0,z}$ in six redshift bins. These distributions are almost Gaussian.}
    \label{FSN}
\end{figure*}

When estimating cosmological parameters, we used the $\chi^2$ statistic for a particular model
with the parameter set of $\theta$ ($H_{0,z_i}$):
\begin{equation}\label{chi2}
\chi_{\theta}^{2}=\chi_{H(z)}^2+\chi_{BAO}^2+\chi_{SNe}^2,
\end{equation}
with 
\begin{equation}
\chi_{H(z)}^{2}=\sum_{i=1}^{N} \frac{\left[H_{\mathrm{obs}}\left(z_{i}\right)-H_{\mathrm{th}}\left(z_{i}\right)\right]^{2}}{\sigma^{2}_i},
\end{equation}
where $H_{obs}(z_i)$ and $\sigma_i$ are the observed Hubble parameter and the corresponding 1$\sigma$ error. Then, $H_{th}(z_i)$ is the value calculated from Equation (\ref{Hz step}). 
\par

The value of $\chi^2_{BAO}$ is 
\begin{equation}
\chi_{BAO}^{2}= \left[\nu_{\mathrm{obs}}\left(z_{i}\right)-\nu_{\mathrm{th}}\left(z_{i}\right)\right]\mathbf{C_{BAO}}^{-1}\left[\nu_{\mathrm{obs}}\left(z_{i}\right)-\nu_{\mathrm{th}}\left(z_{i}\right)\right]^{T},
\end{equation}
where $\nu_{obs}$ is the vector of the BAO measurements at each redshift $z$ (i.e. $D_V\left(r_{s,{\rm fid}}/r_s\right), D_M/r_s, D_H/r_s, D_A/r_s$). The comoving distance $D_M(z)$ is:
\begin{equation}
D_M(z) = (1+z)D_A(z), 
\end{equation}
where $D_A(z)$ is the angular size distance. The radial BAO projection $D_H(z)$ is: 
\begin{equation}
D_H(z) = \frac{c}{H(z)}.     
\end{equation}
The angle-averaged distance $D_V(z)$ is 
\begin{equation}
D_V(z) = \left[czD^2_M(z)/H(z)\right]^{1/3}.  
\end{equation}
The sound horizon of the fiducial model is $r_{s,fid} = 147.5 $ Mpc.

The full covariance matrix of the Pantheon+ SNe Ia sample is considered in the process of calculating $\chi_{SNe}^2$ \citep{2021arXiv211203863S}. This covariance 1701$\times$1701 matrix $\mathbf{C_{SNe}}$ is defined as 
\begin{equation}
\mathbf{C_{SNe}}=\mathbf{D_{stat}}+\mathbf{C_{sys}},
\end{equation}
where $\mathbf{D_{stat}}$ is the statistical matrix and $\mathbf{C_{sys}}$ is the systematic covariance matrix . The value of $\chi^2_{SNe}$ is: 
\begin{equation}
\chi_{SNe}^{2}= \left[\mu_{\mathrm{obs}}\left(z_{i}\right)-\mu_{\mathrm{th}}\left(z_{i}\right)\right]\mathbf{C_{SNe}}^{-1}\left[\mu_{\mathrm{obs}}\left(z_{i}\right)-\mu_{\mathrm{th}}\left(z_{i}\right)\right]^{T}
.\end{equation}
The parameter $\mu_{\mathrm{obs}}(z_i)$ is the distance module from the Pantheon+ sample. The theoretical distance modulus, $\mu_{\textrm{th}}$, is defined as
\begin{equation}
\mu_{th}=5\log_{10} d_{\mathrm{L}} +25,
\label{eq:mu_theory}
\end{equation}
Taking $H_{th}(z)$ from Equation (\ref{Hz step}), the luminosity distance, $d_{L}$, is expressed as
\begin{equation}
d_{L}(z)=c(1+z) \int_{0}^{z} \frac{\mathrm{d} z^{\prime}}{H_{th}(z^{\prime})}.
\end{equation}

The constraints are derived by the Markov Chain Monte Carlo (MCMC) code \textit{emcee} \citep{Foreman-Mackey2013}. The adopted prior is $H_0 \in$ [50,80] $\textrm{km}~\textrm{s}^{-1} \textrm{Mpc}^{-1}$ for all $H_{0,z_i}$. 
Observations of SNe Ia and CMB set tight constraints on the cosmic matter density $\Omega_{m0}$, which are both around $0.3$ \citep{Brout2022,2020A&A...641A...6P}. Thus, a fiducial value $\Omega_{m0}=0.3$ for cosmic matter density was used during the fitting process. We repeated our analysis with the Gaussian prior $\Omega_{m0}=0.315\pm0.007$ from {\it Planck} CMB measurements \citep{2020A&A...641A...6P} and found that it gives very similar results. Thus, the results are largely insensitive to the choice of either prior. 

\subsection{Principal-component analysis of $H_{0,z}$}\label{Sec4}
Although each redshift bin is treated as independent during the fitting process, the results of $H_{0,z_i}$ are still correlated. This is expected, as the integration and summation over low-redshift bins in Equation (\ref{Hz step})
definitely affects the model fit in the middle and higher redshift bins. In order to remove these correlations, we calculate the transformation matrix \citep{2005PhRvD..71b3506H}. The principal component analysis presents a compressed form of the result with all information about the constraint from observations. 

The fitting results impose constraints on the parameters $H_{0,z_i}(i=1...N)$. The covariance matrix can be generated by taking the average over the chain, namely,
\begin{equation}
\mathbf{C}=\langle \mathbf{H} \mathbf{H}^{\rm T}\rangle-\langle\mathbf{H}\rangle\langle\mathbf{H}^{\rm T}\rangle,
\end{equation}
where $\mathbf{H}$ is a vector with components $H_{0,z_i}$ and $\mathbf{H}^T$ is the transpose. We
transformed the covariance matrix to decorrelate the $H_{0,z_i}$ estimations \citep{2005PhRvD..71b3506H}. This was achieved by
finding the uncorrelated basis by diagonalizing the inverse covariance matrix. The Fisher matrix is defined as  
\begin{equation}
\mathbf{F} \equiv \mathbf{C}^{-1} \equiv \mathbf{O}^{\mathrm{T}} {\Lambda} \mathbf{O} ,
\end{equation}
where the matrix $\Lambda$ is the diagonalized covariance for transformed bins. The transformation matrix $\mathbf{T}$ is used as the square root of the Fisher matrix, which is defined as 
\begin{equation}
\mathbf{T}=\mathbf{O}^{\mathrm{T}}  {\Lambda}^{\frac{1}{2}} \mathbf{O} .
\end{equation}
One advantage of this method is that the rows of $\mathbf{T}$ are almost positive across all bands. After being normalized, the rows of $\mathbf{T}$ which means the weights for $H_{0,z_i}$ sum to unity.
Finally, it is clear that the new parameters, 
\begin{equation}
\tilde{\mathbf{H}}=\mathbf{TH,}
\end{equation}
are uncorrelated, because they have the covariance matrix ${\Lambda}^{-1}$. 

After the correlations among $H_{0,z}$ are removed, the corner plots of $H_{0,z}$ are shown in supplementary Figures \ref{H7cor} and \ref{H8cor} for the equal-number binning case and equal-width binning case, respectively. 
The constraints on $H_{0,z}$ are tight and the probability distributions are almost Gaussian.

\begin{figure*}
    \centering
    \includegraphics[width=\textwidth,angle=0]{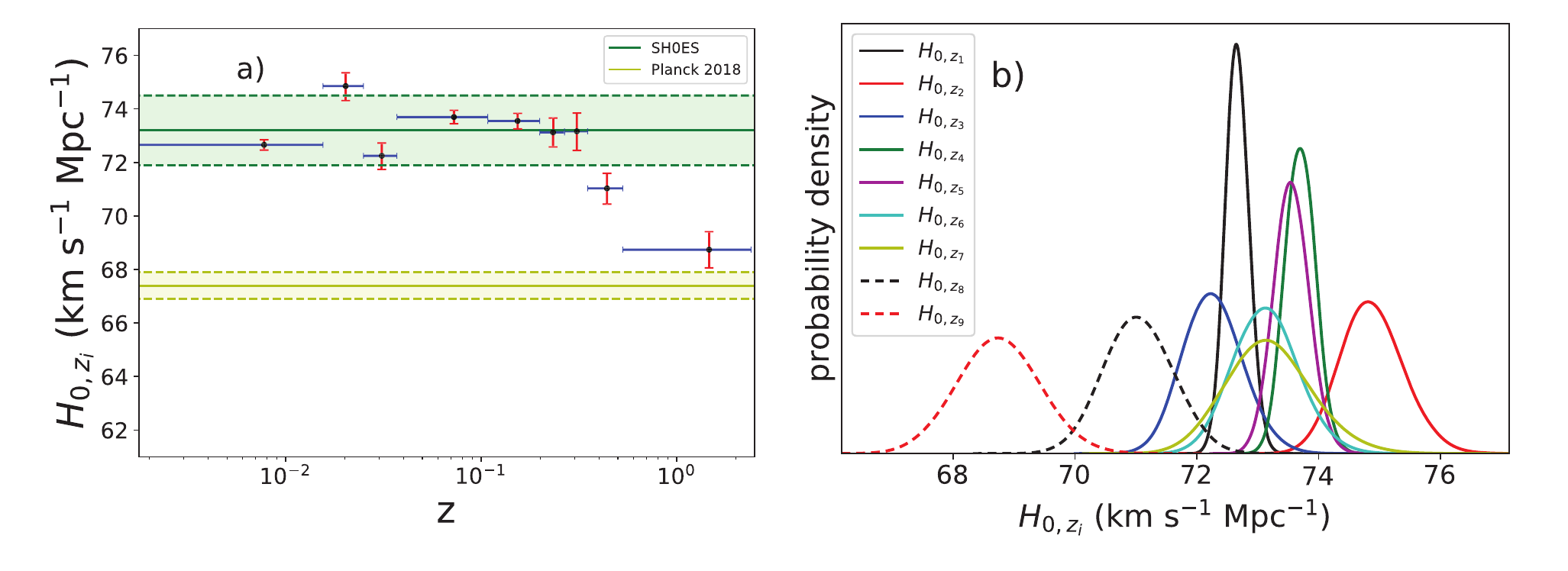}
    \caption{Fitting results of $H_0(z)$ in the equal-number case for nine redshift bins. The left panel shows the value of $H_0(z)$ as a function redshift. There is a clear decreasing trend with $3.8\sigma$ significance at $z>0.3$.  The green line gives $H_{0} = 73.04\pm 1.04$ km s$^{-1}$ Mpc$^{-1}$ from the local distance ladder and its 1$\sigma$ uncertainty \citep{2022ApJ...934L...7R}. The yellow line is the value of $H_{0} = 67.4\pm 0.5$ km s$^{-1}$ Mpc$^{-1}$ from the CMB measurements and its 1 $\sigma$ uncertainty \citep{2020A&A...641A...6P}. The right panel shows the normalized probability distributions of $H_{0,z}$ in nine redshift bins. These distributions are almost Gaussian.}
    \label{Fbin7}
\end{figure*}
    
\section{Results}\label{Sec5}
\subsection{SNe Ia sample}\label{result_SNe}

Before the final result, we estimated the possible evolutionary trend only with the Pantheon+ sample. The whole sample contains 1701 data, most of which are located at low redshift. Therefore, we chose six bins with the upper boundaries of $z_i$ = 0.1, 0.2, 0.3, 0.6, 1.0, 2.4. On account of the relatively few data points in high-redshift bins, some intervals have to be loose. The value of $H_{0,z}$ as a function of redshift (panel a) and the normalized probability distributions (panel b) are shown in Figure \ref{FSN}. The decreasing trend is clear at $z > 0.3$, but the scarce data in high-redshift bins give poor constraints. The 1 $\sigma$ uncertainties for high-redshift bins are so large that we decided to add the observed Hubble parameter data and BAO data \citep{2018ApJ...856....3Y,2022MNRAS.513.5686C}. The joint dataset gives a strict constraint and can be used to research the evolution by different binning methods.

\subsection{Two binning methods}
Two redshift-binning methods were adopted. The first one is equal-number method, namely, the  number of data points in each bin is almost equal \citep{2021ApJ...912..150D}. In contrast, the second one is equal-width method, namely, the bins are equally spaced in redshift \citep{2005PhRvD..71b3506H}. 
For the equal-number method, we chose nine bins with the upper boundaries of $z_i$ = 0.0122, 0.025, 0.037, 0.108, 0.199, 0.267, 0.350, 0.530, and 2.40. Each bin contains about 194 data points (listed in Table \ref{Tbin7results}). More binnings were also performed and we find that the likelihood distributions of $H_{0,z}$ for some bins deviate from Gaussian, due to the scarcity of data points in some high-redshift bins. In the case of
non-Gaussian $H_{0,z}$ distributions, the decorrelation process may introduce some biases. Therefore, nine bins were adopted.

\begin{figure*}
    \centering\includegraphics[width=\textwidth,angle=0]{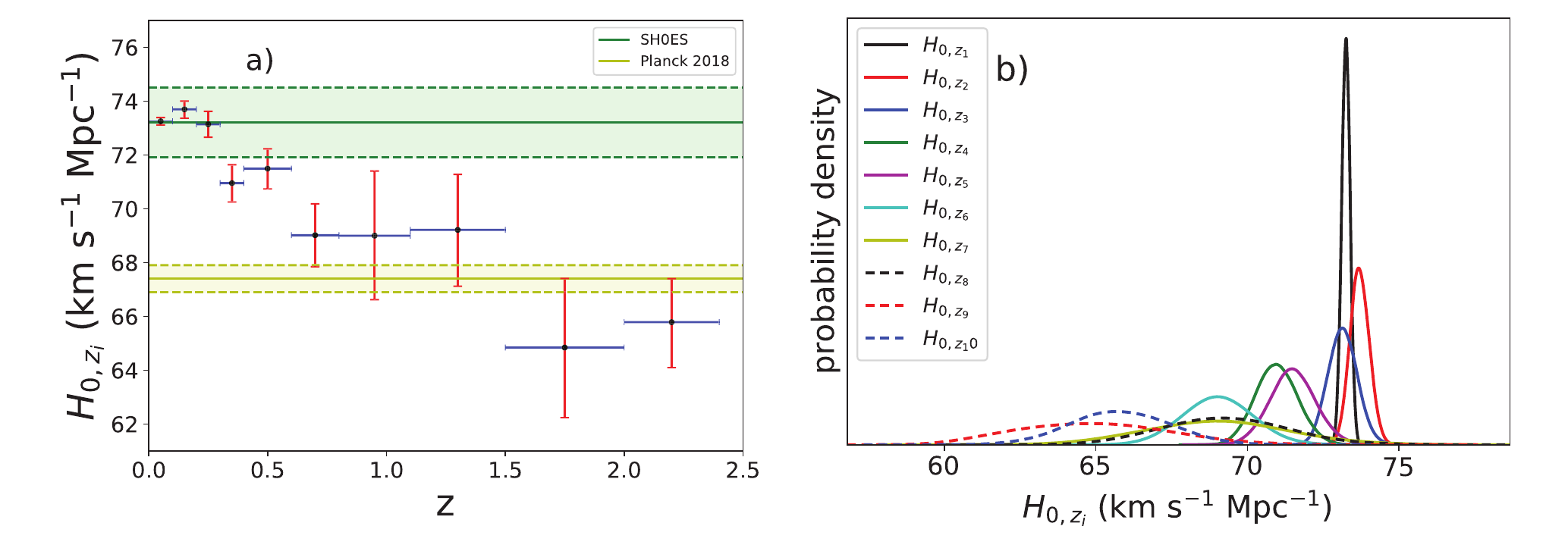}
    \caption{Fitting results of $H_0(z)$ in the equal-width case for ten redshift bins. The left panel shows the value of $H_0(z)$ as a function redshift. There is a clear decreasing trend with $5.6\sigma$ significance at $z>0.3$. The green line gives $H_{0} = 73.04\pm 1.04$ km s$^{-1}$ Mpc$^{-1}$ from local distance ladder and its 1$\sigma$ uncertainty \citep{2022ApJ...934L...7R}. The yellow line is the value of $H_{0} = 67.4\pm 0.5$ km s$^{-1}$ Mpc$^{-1}$ from CMB measurements and its 1$\sigma$ uncertainty \citep{2020A&A...641A...6P}. The right panel shows the normalized probability distributions of $H_{0,z}$ in ten redshift bins. These distributions are almost Gaussian.}
\label{Fbin8}       
\end{figure*}

Altogether, there are nine free parameters in the Markov chain Monte Carlo (MCMC) approach, namely, $H_{0,z_1},...,H_{0,z_9}$. The Pantheon+ SNe Ia sample \citep{2021arXiv211203863S}, BAO data and observed Hubble parameter data were used to estimate the value of $H_{0,z}$ (see Sect. \ref{Sec3}). Given this joint dataset,
we used the MCMC code \textit{emcee} \citep{Foreman-Mackey2013} to sample the parameter set $\theta$ ($H_{0,z_i}$). The prior of $H_0 \in$ [50,80] $\textrm{km}~\textrm{s}^{-1} \textrm{Mpc}^{-1}$ for all $H_{0,z_i}$ was adopted. After removing the correlation among $H_{0,z}$ by diagonalizing the covariance matrix (see Methods), the best-fit results are shown in Table \ref{Tbin7results}. The value of $H_{0,z}$ as a function of redshift (panel a) and the normalized probability distributions (panel b) are shown in Figure \ref{Fbin7}. Due to the large number of data in each bin and a fixed $\Omega_{m0}$, the uncertainty of $H_{0,z}$ is less than 1.0, which is consistent with previous work \citep{2021ApJ...912..150D,2022arXiv220710927W}. In the first seven bins, the fitting results are nearly constant and the value is consistent with the one derived by the local distance ladder \citep{2022ApJ...934L...7R}, however,  it drops rapidly thereafter. It is worth noting that the result in the last bin is consistent with the value from {\it Planck} measurements at a 2$\sigma$ confidence level \citep{2020A&A...641A...6P}. 
\linespread{1}
\begin{table}
        \caption{Fitting results of $H_{0,z_i}$ (in units of $\textrm{km}~\textrm{s}^{-1} \textrm{Mpc}^{-1}$) in the equal-number case. \label{Tbin7results}}
        \centering
        \begin{tabular}{ccc} 
                \\
                \hline
                Redshift bin  & Number (SNe + $H(z)$) &$H_{0,z_i}$   \\
                \hline
                $[0,0.0155]$     & 189  & $72.66^{+0.20}_{-0.19}$ \\
                $[0.0155,0.025]$ & 190  & $74.85^{+0.54}_{-0.50}$ \\
                $[0.025,0.037]$  & 184  & $72.25^{+0.51}_{-0.48}$ \\
                $[0.037,0.108]$  & 193  & $73.70^{+0.25}_{-0.25}$ \\
                $[0.108,0.199]$  & 194  & $73.55^{+0.25}_{-0.25}$ \\
                $[0.199,0.267]$  & 193  & $73.12^{+0.54}_{-0.54}$ \\
                $[0.267,0.350]$  & 195  & $73.17^{+0.72}_{-0.68}$ \\
        $[0.350,0.530]$  & 203  & $71.03^{+0.59}_{-0.56}$ \\
        $[0.530,2.400]$  & 205  & $68.74^{+0.68}_{-0.67}$ \\
                \hline
        \end{tabular}~~~~~~~~
\end{table}

There is an apparent decreasing trend in $H_0(z)$ as a function of redshift. To quantify the significance of this trend, we used the null hypothesis method \citep{Wong2020,Millon2020}. The hypothesis in this situation posits that the values of $H_0(z)$ in each redshift bin are consistent with each other. We first fit a linear regression through each redshift bin. Next, we generated sets of nine mock $H_0$ values with their own uncertainty probability distribution centered around the value of $H_0 = 73.04\pm 1.04$ km s$^{-1}$ Mpc$^{-1}$ \citep{2022ApJ...934L...7R}. The weight of each mock value was 
calculated as follows \citep{Wong2020}. First, the uncertainties’ probability distributions are rescaled so that their maximal values are equal to 1. Then, the area under each distribution is calculated and  the areas are rescaled by their median. Last, the inverse square of the rescaled areas is taken as the weight for each mock value. We also fit a linear regression through the mock value. The slope of the data falls 3.8$\sigma$ away from the mock slope distribution. In other words, the decreasing trend in $H_0(z)$ with increasing redshift has a significance of 3.8$\sigma$.

\linespread{1}
\begin{table}
        \caption{ Fitting results of $H_{0,z_i}$ (in units of $\textrm{km}~\textrm{s}^{-1} \textrm{Mpc}^{-1}$) in the equal-width case. \label{Tbin8results}}
        \centering
        \begin{tabular}{ccc}
                \\
                \hline
                Redshift bin  & Number (SNe + $H(z)$) & $H_{0,z_i}$   \\
                \hline
                $[0,0.1]$   & 743 & $73.25^{+0.14}_{-0.14}$ \\
                $[0.1,0.2]$ & 212 & $73.69^{+0.33}_{-0.31}$ \\
                $[0.2,0.3]$ & 262 & $73.14^{+0.48}_{-0.48}$ \\
                $[0.3,0.4]$ & 190 & $70.95^{+0.70}_{-0.68}$ \\
                $[0.4,0.6]$ & 189 & $71.49^{+0.76}_{-0.74}$ \\
                $[0.6,0.8]$ & 104 & $69.02^{+1.17}_{-1.17}$ \\
                $[0.8,1.1]$ & 16 & $69.00^{+2.38}_{-2.40}$ \\
                $[1.1,1.5]$ & 18 & $69.21^{+2.09}_{-2.06}$ \\
        $[1.5,2.0]$ & 9 & $64.84^{+2.60}_{-2.57}$ \\
        $[2.0,2.4]$ & 3 & $65.78^{+1.69}_{-1.62}$ \\
                \hline
        \end{tabular}
\end{table}

\par
\begin{figure*}
        \includegraphics[width=1\textwidth,angle=0]{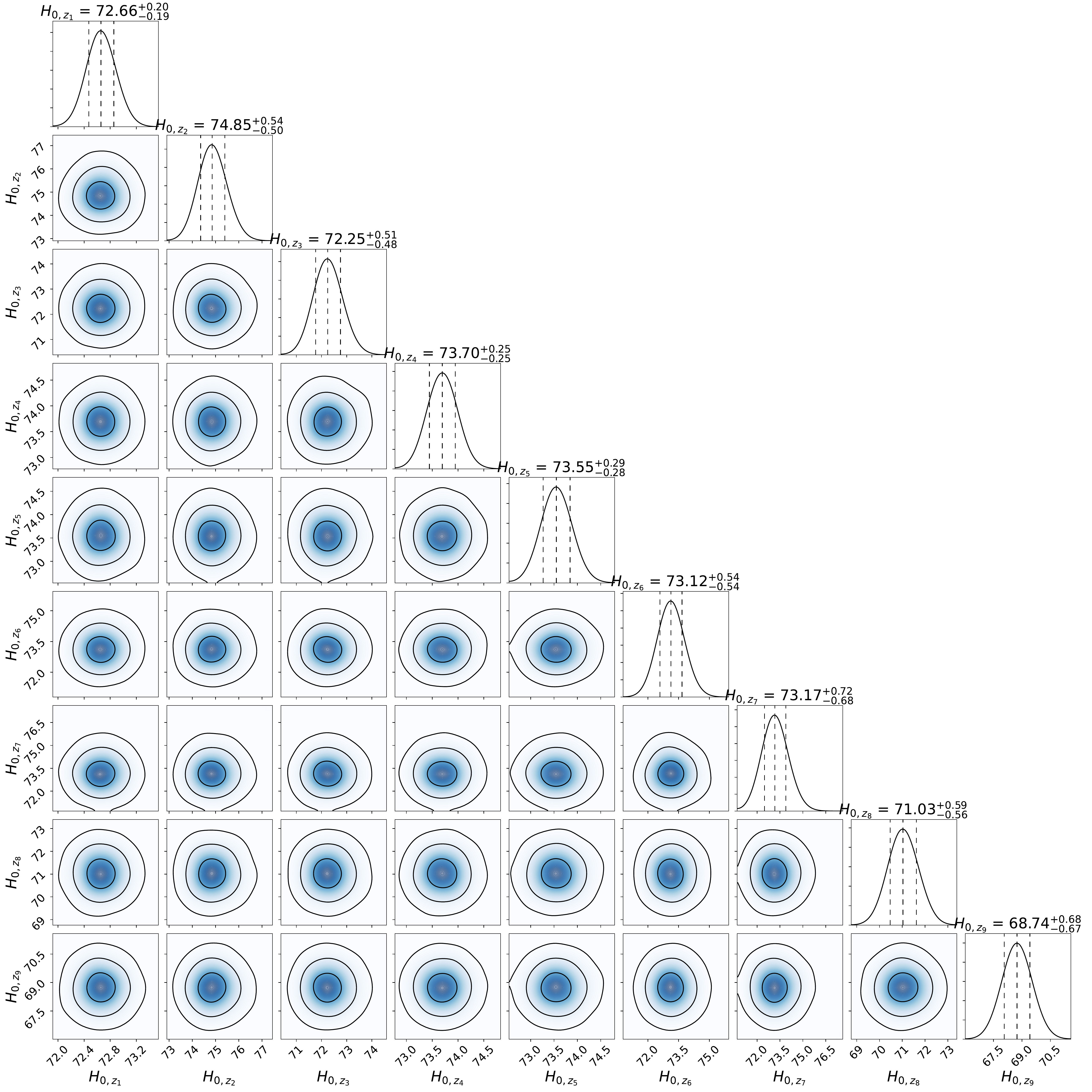}
        \caption{Corner plot of $H_{0,z}$ values from the MCMC code for the equal-number binning case. The panels on the diagonal show the 1D histogram for each parameter obtained by marginalizing over the other parameters. These distributions are almost Gaussian. The off-diagonal panels show two-dimensional projections of the posterior probability distributions for each pair of parameters, with contours to indicate 1$\sigma$ to 3$\sigma$ confidence levels.}
        \label{H7cor}       
\end{figure*}

\begin{figure*}
        \centering
        \includegraphics[width=1\textwidth,angle=0]{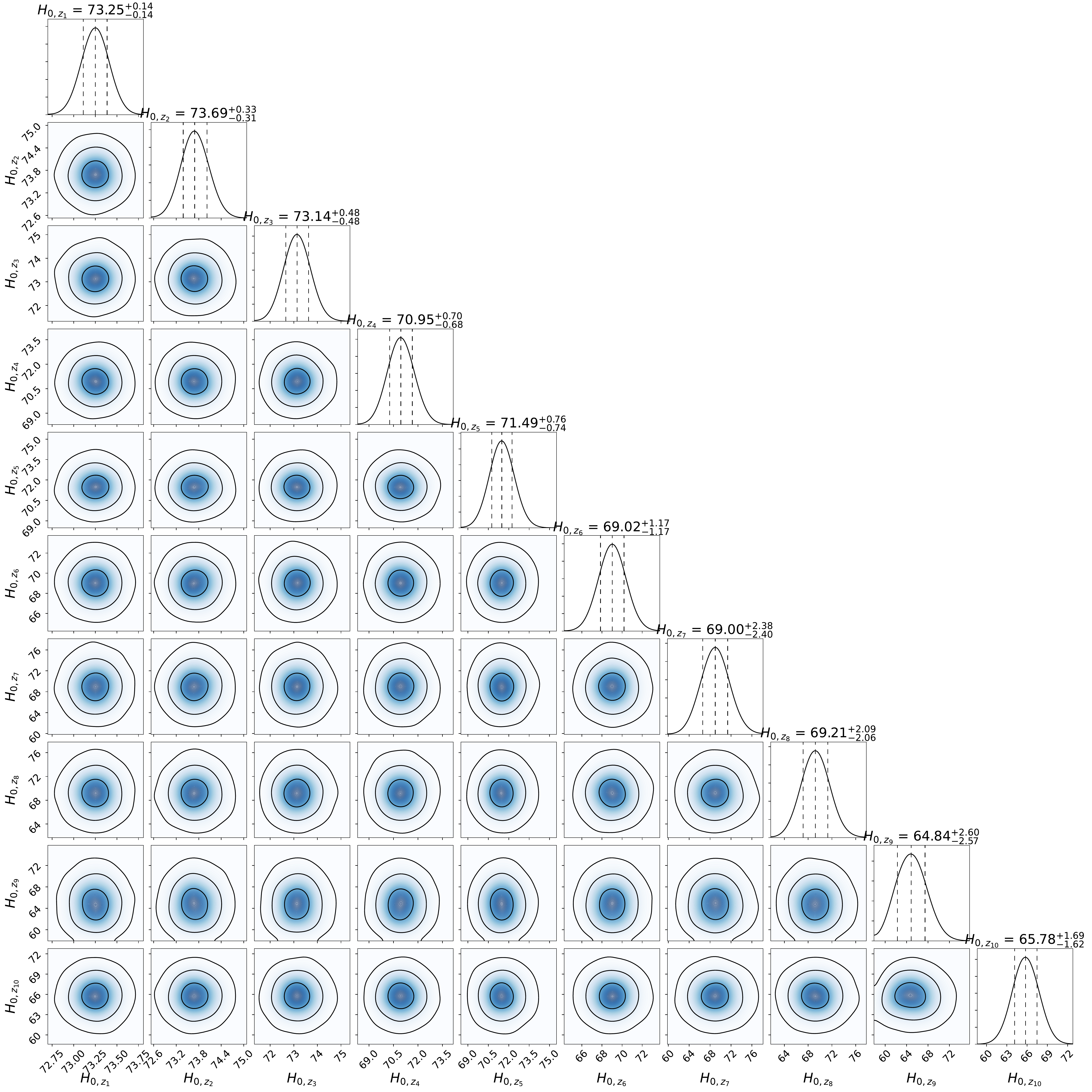}
        \caption{
        Corner plot of $H_{0,z}$ values from the MCMC code for the equal-width binning case. The panels on the diagonal show the 1D histogram for each parameter obtained by marginalizing over the other parameters. These distributions are almost Gaussian. The off-diagonal panels show two-dimensional projections of the posterior probability distributions for each pair of parameters, with contours to indicate 1$\sigma$ to 3$\sigma$ confidence levels. }
        \label{H8cor}       
\end{figure*}

In the second case, the bins are equally spaced with a redshift-width 0.1 in the redshift range [0, 0.4]. Because there are so few data points from $z=0.4$ to $z=2.4$, wider intervals should be adopted in this range. We also tried to equally bin $z$ with a width of $0.1$ in the redshift range [0, 1.0]. The poor constraints in some intervals lead to biases in the decorrelation process. Finally, ten bins are adopted with the upper boundaries $z_i$ = 0.10, 0.20, 0.30, 0.40, 0.6, 0.8, 1.1, 1.5, 2.0, and 2.4. The number of data points and best-fit results are given in Table \ref{Tbin8results}. The uncertainties of the constraints increase gradually as the number of data decreases. The value of $H_{0,z}$ as a function of redshift (panel a) and the normalized probability distributions of $H_{0,z}$ (panel b) are given in Figure \ref{Fbin8}. The fitting results of the first three bins are consistent with the value from the local distance ladder within 1$\sigma$ confidence level \citep{2022ApJ...934L...7R}, and the last two bins are consistent with the value from {\it Planck} CMB measurements within 1$\sigma$ confidence level \citep{2020A&A...641A...6P}. There is a gradually decreasing trend in $H_0(z)$ from the third bin to the last bin. Using the same method as above, the significance of the decreasing trend is 5.6$\sigma$. More importantly, the decreasing trend begins at the same redshift $z\sim 0.3$ for the two binning methods.

Finally, we tested whether additional parameters
to describe $H_0(z)$ are actually needed to improve on a flat $\Lambda$CDM
model fit to the data. For the flat $\Lambda$CDM model, the same value of $\Omega_{m0}=0.3$ was adopted and $H_0$ was left as a free parameter. Two
kinds of standard information criteria were considered: Akaike information criterion (AIC) \citep{Akaike1974} and the Bayesian information criterion (BIC) \citep{Schwarz1974}. Their definitions are: $\mathrm{AIC}=-2\ln L+2k, ~\mathrm{and} ~~ \mathrm{BIC}=-2\ln L+k\ln n,$
where $L\propto e^{-\chi^2/2}$ is the value of the maximum likelihood function, $k$ is the number of model parameters, and $n=1746$ is
the total number of data points. From Table \ref{AICBIC}, there is a significant improvement for both binning methods relative to the flat $\Lambda$CDM model, with $\Delta$AIC of $-44.49$ and $-37.41$, $\Delta$BIC
of $-0.77$ and $-11.77$ for equal-number and equal-width binning methods, respectively. Thus, these values are sufficient to favor the $H_0(z)$ model over $\Lambda$CDM.


\linespread{1.15}
\begin{table}
        \caption{Model comparison \label{AICBIC}}
        \centering
        \begin{tabular}{ccccc} 
                \\
                \hline
                Model & AIC & $\Delta$AIC & BIC & $\Delta$BIC  \\
                \hline
                $\Lambda$CDM  & 1938.08 & 0 & 1943.55 & 0\\
                Equal-number model & 1893.59 & -44.49 & 1942.78 & -0.77\\
                Equal-width model  & 1900.67 & -37.41 & 1955.32 & -11.77\\
                \hline
        \end{tabular}~~~~~~~~
\end{table}


\section{Conclusions }\label{Sec6}
We constrained $H_{0,z}$, defined as the value of $H_0$ derived from the cosmic observations at redshift $z$, and its variation as a function of redshift using a non-parametric
approach. The correlations among $H_{0,z}$ were removed by diagonalizing the covariance matrix. A decreasing trend in $H_{0,z}$ with a significance of 3.8$\sigma$ and 5.6$\sigma$ was found for equal-number and equal-width binning methods, respectively. At low redshift, the value of $H_{0,z}$ is consistent with the value from the local distance ladder and it decreases to the value from CMB measurements at high redshift.
The evolution of $H_{0,z}$ can effectively relieve the Hubble tension without modifications of  early Universe physics. The descending trend may be a signal for the flat $\Lambda$CDM model is breaking down \citep{2021CQGra..38r4001K,2022PhRvD.105f3514K}

The  decreasing behavior found for the Hubble constant with the redshift is significant and urgently calls for an explanation.
At present, the physical mechanism behind the decreasing trend in $H_{0,z}$ is unclear. Two possible origins are suggested by the systematic uncertainties in the data and modifications of the standard cosmological model. 
For the systematic uncertainties, it has been found that the light-curve parameters of SNe Ia, such as the stretch and the color, show no clear dependence on the redshift for the Pantheon SNe Ia sample \citep{Scolnic2018}. However, a recent study found that the SNe Ia SALT2.4 light-curve stretch distribution evolves as a function of redshift \citep{Nicolas2021}, which will affect the value of Hubble constant derived from SNe Ia. Thus, the redshift-dependence of SNe Ia parameters should be extensively studied. If it is not due to selection effects or systematic uncertainties
in the data, our results should be interpreted with physical models. This might indicate the emergence of new physics \citep{2021CQGra..38o3001D,2021A&ARv..29....9S}, such as dynamical dark energy \citep{2017NatAs...1..627Z} or modified gravity models \citep{Kazantzidis2020,2021ApJ...912..150D}. From the Pantheon+ SNe Ia sample, marginal evidence of an increase of cosmic matter density, $\Omega_{m0}$, with a minimum redshift was discovered \citep{Brout2022}, which supports the decreasing trend in the Hubble constant found in this paper. Moreover, a dynamical dark energy signal with $2\sigma$ confidence level was found from Pantheon+ sample \citep{Wang2022}. Due to the lack of high-redshift data, the redshift bins are sparse at $z>0.5$ and $H_{0,z}$ can only be measured below $z=2.4$. In the future, constraints placed on $H_{0,z}$ will improve significantly with upcoming cosmological observations, such as the James Webb Space Telescope, Large Synoptic Survey Telescope, and Euclid and Nancy Grace Roman Space Telescope. In particular, gamma-ray bursts and fast radio bursts may shed light on the evolution of $H_{0,z}$ \citep{Wang2015,Khadka2020,WangF2022,Cao2022,Liang2022,Dainotti2022,Luongo2023,Wu2022}.

\section*{Acknowledgements}
We appreciate the referee for valuable comments and suggestions,
which have helped to improve this manuscript. This work was supported by the National Natural Science
Foundation of China (grant No. 12273009), the China Manned Spaced Project (CMS-CSST-2021-A12), and the Jiangsu Funding Program for Excellent Postdoctoral Talent (20220ZB59).
The numerical code can be found
in the GitHub repository (https://github.com/JoJo20221003/Hz-Code).

\bibliographystyle{aa} 
\bibliography{ref} 

\end{document}